\documentclass[12pt,oneside]{article}
\usepackage{epsfig,  amssymb}
\title{\large\textbf{$^{3}_{\eta}$He nucleus modeling in the frame optical potential model}}
\author{\normalsize{V.A. Tryasuchev\footnote {Electronic address: trs@npi.tpu.ru}, A.V. Isaev}}
\date {\textsl{\small Tomsk Polytechnic University, Tomsk, Russia}}
\begin{document}
\maketitle
\begin{abstract}

The conditions, at which quasi-bound $\eta-^{3}$He state is
possible, have been investigated and compared with the available
findings about $\eta$N-scattering length and the information about
$^{3}$He nucleus from references. We conclude that the existence of
$\eta-^{3}$He quasi-bound state within the framework of the optical
potential model, which doesn't contradict all collected findings, is
not possible, but the observing anomaly of $\eta-^{3}$He-interaction
at low energies is a virtual state.

\textsl{PACS}: 21.10.-k
\end{abstract}

Quark-compound meson having the final lifetime can form short-living
bound states with atom nuclei called mesic nuclei in case of it
interaction of attraction with nucleons. Analyzing the result of the
precise experiment [1], the authors of the work [2] have shown that
the quasi-bound state existence of $\eta-^{3}$He or the presence
of virtual state of the system is possible. Considering that it has
been $\eta$-nucleus of $\eta-^{3}$He (quasi-bound state), we have
investigated the conditions in the frame optical potential model, at
which its formation is possible. The obtained conditions have been
compared with the available findings about $\eta$N-scattering length
and the information about $^{3}$He nucleus, which is in the
literature.

The optical potential for $\eta$-nucleus interaction is written in
the following form:
\begin{equation}
    2U(r)\mu = -4\pi(1+\frac{m_{\eta}}{m_{N}})\rho (r)a_{0}
\end{equation}
where, $m_{\eta}$, $m_{N}$ are meson and nucleon masses, $\mu$ is
reduced meson-nucleus mass, $a_{0}$ is $\eta$N-scattering length,
$\rho(r)$ is the spherically symmetrical nucleon density of nucleus,
which has been chosen in Fermi form:
\begin{equation}
    \rho (r)=\frac{\rho _{0}}{1+\exp (\frac{r-R_{c}}{a})}
\end{equation}
where $R_{c}$ is the half-density radius, $a$ is the diffusion
thickness of nucleus, $\rho_{0}$ is the maximum nucleon density of
nucleus. For $^{3}$He, $R_{c}$ and $a$ can be determined from the
following conditions:
\begin{equation}
3 = \int^{\infty}_{0}r^{2}\rho (r)dr,
\left\langle r^{2} \right\rangle = \frac{1}{3}\int^{\infty}_{0}r^{4}\rho (r)dr
\end{equation}
where $\left\langle r^{2}\right\rangle^{1/2}$  is the
root-mean-square radius of $^{3}$He nucleus determined during the
experiment, which has been taken as being equal to 1.9 Fm, as the
mean from some works, where this radius has been measured or
determined. The obtained distributions of nucleon density of nucleus
$^{3}$He at different diffusion $a/R_{c}$ is shown in figure 1.

\begin{figure}[h]
    \centering
        \includegraphics[width=10cm,height=8cm]{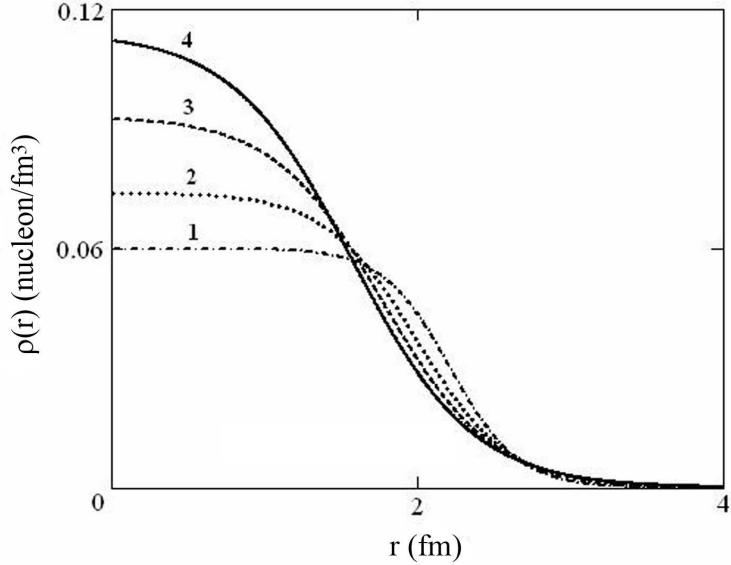}
   \label{Fig.1}
 \caption{The distribution of nucleon density of nucleus $^{3}$He for
different values of diffusion  ${a}/R_{c}$. The curve
parameters are given in table 1.}
\end{figure}

\begin{table}[ht]
\centering
\caption{}
\begin{tabular}{|cccc|}\hline\hline
$N^{0}_{-}$ & $a/R_{c}$ & $R_{c}$, Fm & $\rho_{0}$, nucleon/Fm$^{3}$\\\hline
1 & 0.10 & 2.210 & 0.060\\
2 & 0.15 & 1.991 & 0.074\\
3 & 0.20 & 1.770 & 0.093\\
4 & 0.25 & 1.571 & 0.114\\\hline\hline
\end{tabular}
\end{table}

As it is shown in the figure, for the diffusion, which is more then
0.2, the nucleon density distribution in nucleus $^{3}$He doesn't
have plateau in the middle, that, in our opinion, corresponds to the
reality, so we have stopped at these values ${a}/R_{c}$.

For quasi-bound state formation in the complex potential with own
complex energy $E=-(\epsilon+i\frac{\Gamma}{2})$ where $\epsilon$ is
the binding energy, and $\Gamma$ is the level width, the definite
relation between the absolute values of imaginary and real parts of
potential [3,4] is required. We have calculated the formation
boundary of $\epsilon\approx0$ of the discussed $\eta$-nucleus in
the dependence of imaginary potential part on the real one at
different nucleon densities of nucleus $^{3}$He. This dependence has
been shown in the complex plane of free $\eta$N-scattering length
(fig.2).

\begin{figure}[ht]
    \centering
        \includegraphics[width=10cm,height=8cm]{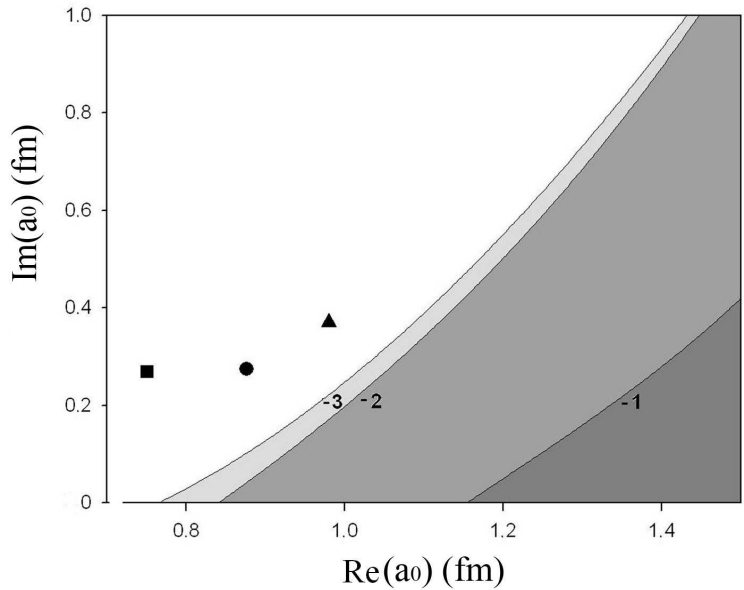}
    \caption{Curves are the formation boundaries of quasi-bounded states
for $\eta-^{3}$He, the darkened areas are the existence areas of
quasi-bounded states for $\eta$N-scattering length with different
parameters of $^{3}$He nucleus diffuseness: $1 - 0.1$, $2 - 0.15$,
$3 - 0.25$, in the complex plane; $\eta$N scattering lengths of
interaction have been taken from works: \small $\blacksquare$
\normalsize $-$ $[6]$, \Large $\bullet$ \normalsize $-$ $[7]$,
\normalsize $\blacktriangle$ $-$ $[8]$.}
    \label{fig:fig2}
\end{figure}

 It is evident from this figure, that the
quasi-bound state occurrence of system $\eta-^{3}$He with the low
binding energy is possible at $\left|Re(a_{0})\right|\geq0.75$ Fm
for the values of nucleus $^{3}$He diffusion, $a/R_{c}=0.25$. It
should be noticed that at the further diffusion increasing,
$a/R_{c}<0.25$, the boundary of $\eta$-mesic nucleus formation
practically doesn't shift. The obtained limit for $\eta$N-scattering
length from the existence of $^{3}_{\eta}$He doesn't contradict the
available quantity values of $\left|Re(a_{0})\right|$ in set works,
which are in the interval $(0.27\div1.0)$ Fm (see work [5]). But
their imaginary parts don't allow the $^{3}_{\eta}$He formation to
be considered as possible.

Thus, is the interaction of $\eta$-meson with the nucleus is
considered in the optical potential model, so the formation of
quasi-bound state $\eta-^{3}$He is not possible (that is, nucleus
$^{3}_{\eta}$He), because its existence would contradict the obtained
knowledge about $\eta$N-scattering length and about the nucleon
density distribution of nucleus $^{3}$He. It follows from our
calculations, that in the experiment [1] the virtual state
$\eta-^{3}$He, which is close by energy to the sum mass of free
particles, has been revealed. Off-shell $\eta$N-scattering length
use in potential (1), as the authors of work [5] insist,
couldn't change our conclusion, because the quantity decreasing of
real and imaginary parts $a_{0}$ in this case must occur
proportionally. At the same time our result agrees to the work [9]
conclusion about the impossibility of the existence of quasi-bounded
states of $\eta$-meson with three nucleons at different
$\eta$N-scattering lengths, which are in the special literature,
while the formation of virtual state $\eta-^{3}$He is possible
[9].

\end{document}